# Optimizing System Memory Bandwidth with Micron CXL<sup>TM</sup> Memory Expansion Modules on Intel® Xeon® 6 Processors

Rohit Sehgal, Vishal Tanna
Micron Technology
San Jose, CA

Vinicius Petrucci
Micron Technology
Austin, TX

Anil Godbole
Intel Corporation
Santa Clara, CA

*Abstract*— High-Performance Computing (HPC) and Artificial Intelligence (AI) workloads typically demand substantial memory bandwidth and, to a degree, memory capacity. CXL<sup>TM</sup> memory expansion modules, also known as CXL "type-3" devices, enable enhancements in both memory capacity and bandwidth for server systems by utilizing the CXL protocol which runs over the PCIe interfaces of the processor. This paper discusses experimental findings on achieving increased memory bandwidth for HPC and AI workloads using Micron's CXL modules. This is the first study that presents real data experiments utilizing eight CXL E3.S (x8) Micron CZ122 devices on the Intel® Xeon® 6 processor 6900P (previously codenamed Granite Rapids AP) featuring 128 cores, alongside Micron DDR-5 memory operating at 6400 MT/s on each of the CPU's 12 DRAM channels. The eight CXL memories were set up as a unified NUMA configuration, employing software-based page level interleaving mechanism, available in Linux kernel v6.9+, between DDR5 and CXL memory nodes to improve overall system bandwidth. Memory expansion via CXL boosts read-only bandwidth by 24% and mixed read/write bandwidth by up to 39%. Across HPC and AI workloads, the geometric mean of performance speedups is 24%.

## I. INTRODUCTION

High-performance and AI workloads encompass important computational tasks that demand substantial processing and memory resources. These workloads are frequently utilized in scientific research, simulations, and data-intensive applications, including computational fluid dynamics, weather forecasting, and DNA sequencing.

Alongside HPC, AI plays a crucial role in analyzing large datasets and driving innovations across various fields. For example, LLM inference and vector search in Retrieval-Augmented Generation (RAG) are crucial workloads as they enable efficient access to relevant information and enhance the quality of generated responses, making AI interactions more accurate and contextually aware.

This paper presents experimental work conducted by Micron and Intel, which examines the performance of AI and HPC workloads on the Intel® Xeon® 6 processor 6900P series now in full production, paired with Micron CZ122 CXL devices. The study quantifies the performance benefits of utilizing Micron CZ122 devices in HPC/AI workloads, noting improvements in performance by expanding system memory bandwidth using CXL memory expansion, beyond local DRAM modules. The memory bandwidth expansion enabled by CXL is essential for enhancing the performance of HPC and AI workloads.

*While CXL has primarily aimed at expanding memory capacity, its advantages for bandwidth-intensive workloads still need to be thoroughly explored and quantified in real CXL-capable systems, utilizing as many supported PCIe lanes as possible.* In particular, the unique bandwidth characteristics of local DRAM and CXL memory can differ depending on the read/write ratio of workloads, creating challenges in optimizing the capabilities of each memory tier in terms of memory bandwidth. For this purpose, a software-based weighted interleaving method, available in mainstream Linux kernel distribution, is employed for optimization.

## II. PLATFORM CONFIGURATION

### A. Intel Xeon 6 CPU System (Avenue City platform)

The 6900P CPU supports 6 x16 (96) PCIe 5.0 lanes. The lanes support CXL 2.0 Type-3 devices, allowing for memory expansion. The CPU supports any four x16 lanes to be used as CXL links.

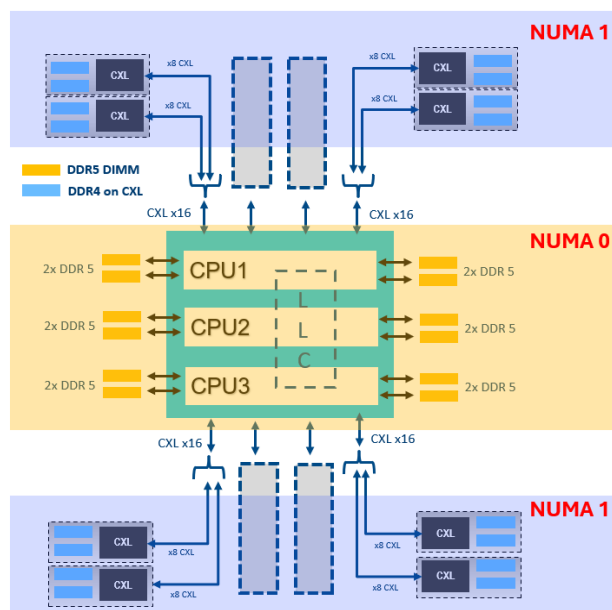

*Figure 1. System Architecture of Intel Xeon 6 processor 6900P with 128 cores and 12x Micron DDR5 6400 MT/s. All 12 local DRAM channels are designated as NUMA node 0 (HEX mode), while all the Micron's CXL modules (8 in total) are brought up as separate NUMA node 1.*

As the focus of this paper is on demonstrating the effectiveness of increasing bandwidth rather than capacity, smaller memory modules were intentionally chosen for both native DRAM (64 GB) and CXL (128 GB) modules.

The system configuration employed (Figure 1) facilitates the management of various memory tiers by efficiently organizing and distinguishing between the locally attached DRAM and the CXL memory modules.

Traditionally, the Linux kernel has managed memory allocation across multiple NUMA (Non-Uniform Memory Access) nodes. Each of the memory types (either DRAM or CXL) is represented as a single NUMA node, allowing the system to use existing abstractions to manage and allocate memory across these two different pools.

Recently, NUMA nodes have been used to categorize memory into performance tiers, while existing allocation policies can place memory on specific NUMA nodes. For example, when brought up as system memory, CXL memory is treated as a separate NUMA node.

To showcase the advantages of using CXL memories, the system configuration is designed so that the local RDIMM slots are filled with the fastest available Micron RDIMMs, delivering a bandwidth of 6400 MT/s per slot. All 12 available slots are populated – totaling 768GB memory capacity. As shown in Figure 2, eight Micron CZ122 128GB CXL devices are utilized, occupying 64 PCIe lanes and providing a total additional memory capacity of 1TB.

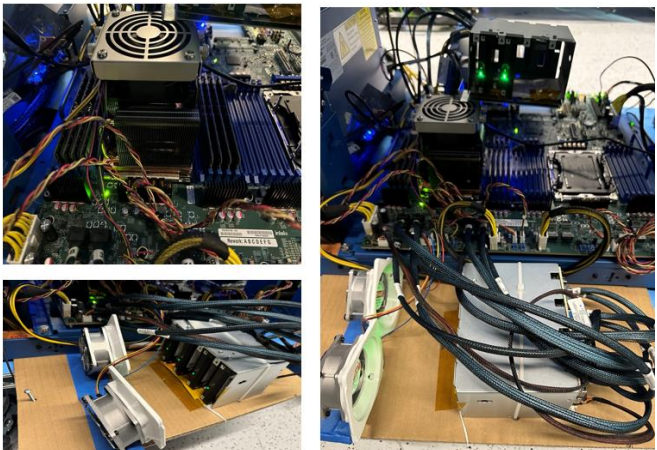

*Figure 2. Configuration of Micron CZ122 CXL modules with an Intel Xeon-6 CPU on an Avenue City platform involves connecting four cards directly to the backplane, while the other four cards are attached using riser cards in two CME slots.*

Additional details on the platform are shown in the table below.

| Platform | Intel Avenue City |
|---|---|
| **CPU family** | Intel® Xeon® 6 6900P series with 128 cores |
| **Native DRAM** | Micron DDR5-64GB (6400MTs) (12 modules ~ 768 GB) – HEX mode |
| **CXL Memory** | Micron CZ122 – 128GB * 8 (8 modules E3.S form factor ~ 1TB) |

| OS | Red Hat Enterprise Linux 9.4 |
|---|---|
| **Kernel** | 6.11.6 (With support for weighted memory interleaving) |

### B. Memory Expansion with Micron CZ122 CXL modules

Micron's CZ122 CXL modules are currently in production and have demonstrated reliable performance across various workloads, effectively showcasing memory expansion over CXL interface. The addition of these CXL modules enhance both the memory bandwidth and the capacity of the server, building on what is already provided by the RDIMM slots; that is, delivering memory bandwidth expansion.

Optimally placing newly allocated pages is a complex issue. NUMA interleaving, a traditional approach under Linux, evenly distributes pages across memory nodes for consistent performance. However, it lacks the ability to consider memory tier performance differences.

A recent series of patches has added weighted NUMA interleaving capabilities to the Linux kernel, allowing for more strategic memory allocation based on performance characteristics of different memory nodes in system. This strategy optimizes system memory bandwidth by effectively utilizing bandwidth both local DRAM and CXL memory nodes. The weighted-interleaving feature, introduced in Linux kernel version 6.9+ and influenced significantly by Micron's contributions, enables the adjustment of weights assigned to individual pages across various memory types, thereby enhancing overall memory bandwidth (as illustrated in Figure 2).

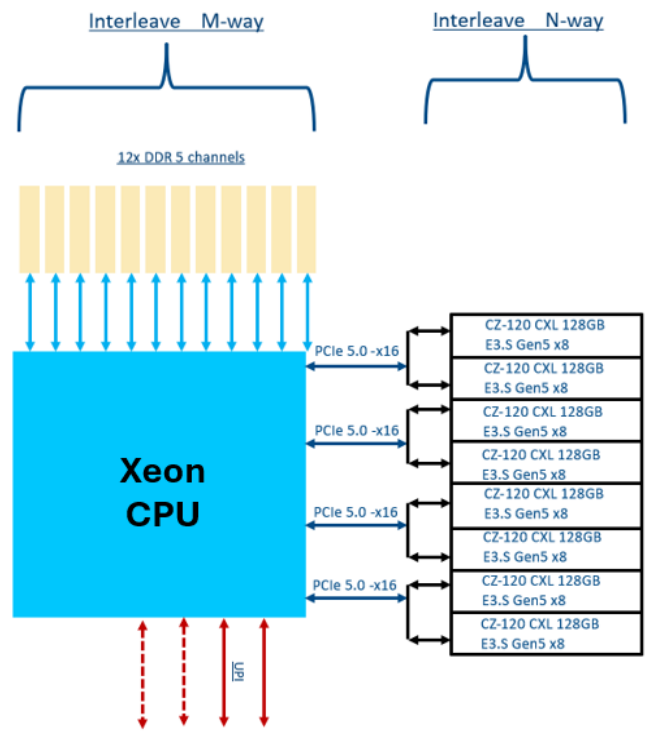

*Figure 3. Software-based weighted interleaving (M:N) allowing placing M pages on local DRAM and N pages on CXL memory for optimized system memory bandwidth.*

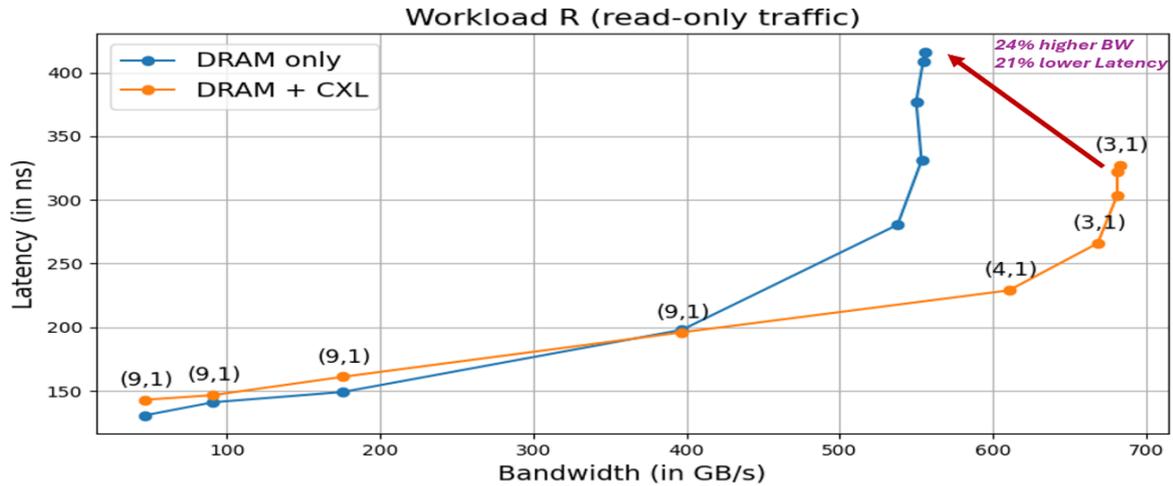

*Figure 4. Bandwidth vs Latency curves using DRAM only vs DRAM + CXL. The interleaving weights are represented as pairs (DRAM, CXL). It's important to note that at low bandwidth, a greater number of pages (9) are allocated to DRAM compared to CXL (1), as indicated by the weights (9,1). Conversely, under high load conditions, the optimal interleaving weights shift to (3,1).*

### III. NATIVE DRAM VS. CXL ATTACHED MEMORY PERFORMANCE CHARACTERISTICS

Before the performance analysis of the actual workloads is introduced, the performance characteristics of local DRAM and CXL memory regarding bandwidth at various read-to-write ratios of memory traffic will be presented and discussed[1].

| Workload | Memory Tier | Bandwidth (in GB/s) | Bandwidth (Normalized) | CXL over DRAM (Theoretical gains with CXL) |
|---|---|---|---|---|
| Read only | DRAM | 556 | 1.00 | - |
| 3R,1W | DRAM | 486 | 0.87 | - |
| 2R,1W | DRAM | 474 | 0.85 | - |
| 2R,1W (non-temporal W) | DRAM | 466 | 0.84 | - |
| 1R,1W | DRAM | 446 | 0.80 | - |
| Read only | CXL | 205 | 1.00 | **37%** |
| 3R,1W | CXL | 214 | 1.04 | **44%** |
| 2R,1W | CXL | 208 | 1.01 | **44%** |
| 2R,1W (non-temporal W) | CXL | 189 | 0.92 | **41%** |
| 1R,1W | CXL | 214 | 1.04 | **48%** |

The performance data from table above indicates that DRAM performs optimally in read-only workloads, but its performance diminishes when the number of writes is equal to or exceeds the number of reads. For instance, in a workload with a 1:1 read to write ratio, DRAM's performance drops by 20% compared to a read-only scenario.

Conversely, CXL memory demonstrates the opposite trend due to the bidirectional nature of the PCIe interface, resulting in better performance for mixed read-write workloads. Another noteworthy observation is that CXL memory shows an 8% decrease in bandwidth during a non-temporal write workload. Therefore, it's crucial to analyze the read-to-write ratio of a workload to identify the optimal interleaving strategy for utilizing DRAM and CXL memory tiers effectively.

As shown in Figure 4, it's also important to note that memory latency is reduced when using CXL. This is because workloads that rely solely on local DRAM can be bandwidth-limited, leading to significantly higher memory access latency (loaded latency) under heavy loads. In contrast, combining DRAM with CXL memory through optimized weighted interleaving results in lower latency, despite CXL memory having a higher unloaded latency.

At each data point on the "DRAM + CXL" curve, the interleave ratio of DRAM and CXL is displayed. Under low bandwidth conditions, it's advantageous to utilize more DRAM due to its lower latency compared to CXL memory (9:1 ratio). However, as the load increases, the reliance on DRAM decreases while the emphasis shifts towards CXL memory. Ultimately, a 3:1 ratio was identified as optimal under maximum load conditions for a read-only workload traffic.

As will be discussed in the next sections, evaluating the use of CXL memory alongside local DRAM reveals various performance improvements. For instance, in a read-only scenario (where DRAM excels) the addition of CXL memory bandwidth results in a **24%** performance boost. The upcoming experiments will demonstrate that for mixed read/write workloads, the performance improvements with CXL, attributed to balanced memory interleaving, can reach as high as **39%**. The following sections will demonstrate that for different workload mixes, we may need to adjust the interleaving weights based on the read-to-write ratio of the workload.

---

[1] Performance results are derived from testing in the specified configuration (Section II.A). Results may vary, so it is recommended to reconfirm them in your setting.

## IV. WORKLOAD ANALYSIS

### A. Intel MLC (Microbenchmark)

Intel MLC (Memory Latency Checker) is a microbenchmark tool designed to assess memory latencies and bandwidth in computer systems. It helps analyze how these metrics change under varying loads, providing insights into the performance of the memory subsystem.

Utilizing the software-based interleaving kernel feature, memory allocation between DRAM and CXL is determined based on a user-defined ratio. Bandwidth measurements are obtained by running the MLC workloads with different read:write ratios.

The weights for each memory tier are given in terms of the *number of pages allocated on DRAM versus CXL memory*. For example, a weight of 3 (DRAM) and weight of 1 (CXL) means 75% of the pages (and eventually the associated memory traffic) allocated on DRAM, while 25% allocated to CXL memory. The following tables presents the results of MLC for various read:write ratios.

**Workload: R (read-only)**

| Weight (DRAM) | Weight (CXL) | BW (in GB/s) | BW (Normalized) |
|---|---|---|---|
| 1 | 0 | 556 | 1.00 |
| 1 | 1 | 394 | 0.71 |
| 2 | 1 | 590 | 1.06 |
| 5 | 2 | 669 | 1.20 |
| **3** | **1** | **690** | **1.24** |
| 4 | 1 | 677 | 1.22 |
| 0 | 1 | 205 | 0.37 |

As shown above, MLC results for the **R (Read-only)** workload indicate a **24%** increase in bandwidth with a 3:1 interleave ratio of DRAM to CXL.

**Workload: W2 (2W, 1R)**

| Weight (DRAM) | Weight (CXL) | BW (in GB/s) | BW (Normalized) |
|---|---|---|---|
| 1 | 0 | 474 | 1.00 |
| 1 | 1 | 422 | 0.89 |
| 2 | 1 | 624 | 1.32 |
| **5** | **2** | **636** | **1.34** |
| 3 | 1 | 617 | 1.30 |
| 4 | 1 | 586 | 1.24 |
| 0 | 1 | 208 | 0.44 |

As shown above, MLC results for the **W2 (2W, 1R)** workload indicate a **34%** increase in bandwidth with a 5:2 interleave ratio of DRAM to CXL.

**Workload: W5 (1W, 1W)**

| Weight (DRAM) | Weight (CXL) | BW (in GB/s) | BW (Normalized) |
|---|---|---|---|
| 1 | 0 | 446 | 1.00 |
| 1 | 1 | 409 | 0.92 |
| **2** | **1** | **621** | **1.39** |
| 5 | 2 | 614 | 1.37 |
| 3 | 1 | 585 | 1.31 |
| 4 | 1 | 551 | 1.24 |
| 0 | 1 | 214 | 0.48 |

As shown above, MLC results for the **W5 (1W, 1R)** workload indicate a **39%** increase in bandwidth with a 5:2 interleave ratio of DRAM to CXL.

**Workload: W10 (2R, 1W non-temporal)**

| Weight (DRAM) | Weight (CXL) | BW (in GB/s) | BW (Normalized) |
|---|---|---|---|
| 1 | 0 | 466 | 1.00 |
| 1 | 1 | 390 | 0.84 |
| 2 | 1 | 533 | 1.14 |
| **5** | **2** | **607** | **1.30** |
| 3 | 1 | 601 | 1.29 |
| 4 | 1 | 572 | 1.23 |
| 0 | 1 | 189 | 0.41 |

As shown above, MLC results for the **W10 (2R, 1W non-temporal)** workload indicate a **30%** increase in bandwidth with a 5:2 interleave ratio of DRAM to CXL.

In summary, as seen in the tables above, for the 100% Read workload, splitting the pages between DRAM and CXL in a 3:1 ratio (3 pages in DRAM, 1 in CXL) results in a 24% bandwidth gain compared to using only DRAM.

For the W2, W3, W5, and W10 MLC workloads, the optimal performance occurs with a DRAM to CXL ratio of 5 to 2. This configuration yields a 34-39% bandwidth increase over DRAM alone. The MLC data shows that adding CXL memory significantly boosts bandwidth.

It is worth noticing that the MLC data provides us with a upper bound on the performance gains when the workload is memory-bandwidth bound given a particular read:write ratio.

For instance, LLM inference predominantly involves read-only traffic, with bottlenecks generally arising at the token generation stage, which necessitates repeated reading of model weights for each token. Consequently, the optimal interleave ratio should be 3:1 for DRAM to CXL memory.

### B. AI Workloads

The Intel Xeon 6 processor with P-cores family is optimized for HPC and AI workloads, enhancing performance in deep learning and machine learning applications. Optimizations take advantage of Intel® Advanced Vector Extensions 512

(Intel® AVX-512) Vector Neural Network Instructions (VNNI) and Intel® Advanced Matrix Extensions (Intel® AMX) on Intel CPUs.

With 128 physical cores, the CPU architecture provides specialized acceleration for AI operations, improving throughput and reducing latency in LLM inferencing and vector search workloads. The architecture supports matrix multiplication and efficiently handles models with billions of parameters.

**LLM Inference -** To run LLM inferencing on the Intel hardware, the open-source Intel Extensions for PyTorch (IPEX) was used. IPEX has up to date optimizations for an extra performance boost on Intel hardware. The LLM model used was Meta-Llama3-8B-Instruct. The data type employed for the weights is 'bfloat16'. Batch size of one was used. With using the intel pytorch extensions for inferencing, the LLAMA3-8B-Instruct gave a speed up of 17% with 3:1 DRAM to CXL ratio versus using DRAM only memory.

| Weight (DRAM) | Weight (CXL) | Output Token Latency (ms) | Speedup |
|---|---|---|---|
| 1 | 0 | 42.91 | 1.00 |
| 2 | 1 | 40.43 | 1.06 |
| 5 | 2 | 37.54 | 1.14 |
| **3** | **1** | **36.83** | **1.17** |

**FAISS (Vector Search) -** FAISS [7] is a library developed by Facebook AI for efficient similarity search and clustering of dense vectors. The dataset used was the Microsoft Turing-ANNS consisting of a raw vector space of one billion points with 100 dimensions, using L2 distance and k-NN method. As recommended by Meta [8], the index used was: OPQ128_256-IVF65536_HNSW32-PQ128x4fsr. This is an optimized FAISS index configuration that specifies a series of transformations and indexing methods for efficient similarity search. Here is a breakdown of what each part means:

- **OPQ128_256**: Optimized Product Quantization rotates vectors for efficient encoding, with 128 and 256 dimensions involved.
- **IVF65536**: Inverted File Index with 65,536 clusters speeds up the search by dividing the vector space into clusters.
- **HNSW32**: Hierarchical Navigable Small World graph with 32 neighbors, a graph-based method for approximate nearest neighbor search.
- **PQ128x4fsr**: Product Quantization with 128 dimensions and 4 subquantizers for further optimizations.

The configuration combines several advanced techniques to create an efficient and scalable index for similarity search in large datasets.

To report the final performance data, these parameters were configured: nprobe=4096 and efSearch=512. Both are crucial for balancing speed and accuracy in FAISS searches. A higher nprobe (number of clusters probed) increases accuracy but also search time. Similarly, efSearch (number of candidate nodes explored) enhances accuracy at the cost of search time. These values were optimized to achieve a high recall rate with minimal search time. The configuration resulted in a recall rate of 77% @ 10, meaning 77% of the true nearest neighbors are included in the top 10 results returned by the search algorithm.

| Weight (DRAM) | Weight (CXL) | Time (ms / query) | Speedup |
|---|---|---|---|
| 1 | 0 | 0.545 | 1.00 |
| **2** | **1** | **0.442** | **1.23** |
| 5 | 2 | 0.454 | 1.20 |

The FAISS workload demonstrated a 23% improvement with a DRAM to CXL ratio of 2:1.

*C. HPC Workloads*

HPC workloads stand for High performance workloads – those include OpenFOAM, HPCG, Xcompact3d, POT3D. These workloads typically require higher memory bandwidths in addition to increased capacity.

**OpenFOAM** - OpenFOAM workload benchmarks are standardized test cases designed to evaluate the performance and scalability of hardware and software configurations when running OpenFOAM, an open-source computational fluid dynamics (CFD) software. These benchmarks simulate various fluid dynamics scenarios to assess how efficiently different systems handle complex CFD computations. The OpenFOAM drivaerFastback case was used with an input of approximately 200 million cells. The results from the benchmark for different DRAM/CXL ratios are shown below:

| Weight (DRAM) | Weight (CXL) | Execution time (s) | Speedup |
|---|---|---|---|
| 1 | 0 | 254 | 1.00 |
| 2 | 1 | 212 | 1.20 |
| **5** | **2** | **209** | **1.22** |
| 3 | 1 | 210 | 1.21 |

The OpenFOAM workload has exhibited a 22% improvement with a DRAM to CXL ratio of 5:2.

**HPCG** - The High-Performance Conjugate Gradients (HPCG) benchmark is a workload designed to assess supercomputing systems by solving a large, sparse linear system using a multigrid preconditioned conjugate gradient algorithm. Unlike the High Performance Linpack (HPL) benchmark, which focuses on dense matrix computations, HPCG emphasizes memory access patterns and data movement, reflecting the behavior of real-world scientific and engineering applications. By doing so, HPCG provides a more comprehensive measure of a system's capability to handle complex, memory-intensive workloads. The input used was the following: x=192, y=192, z=192. Results are shown in the table below.

| Weight (DRAM) | Weight (CXL) | Performance (GFlops/s) | Speedup |
|---|---|---|---|
| 1 | 0 | 92 | 1.00 |
| 2 | 1 | 111 | 1.20 |
| 5 | 2 | 113 | 1.23 |
| **3** | **1** | **117** | **1.27** |

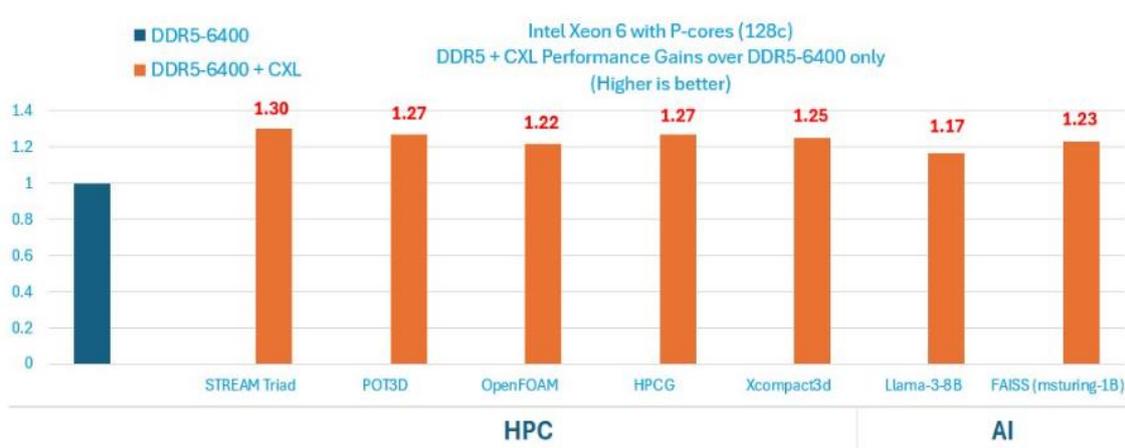

*Figure 5. Summary of performance gains for the HPC and AI workloads running on DDR5-6400 (baseline) vs. DDR5-6400 + CXL.*

The HPCG benchmark has shown 27% improvement with DRAM: CXL = 3:1

**Xcompact3D** - The Xcompact3D benchmark is a performance evaluation tool designed to assess computational efficiency when solving the incompressible Navier-Stokes equations using the Xcompact3D solver. It focuses on simulating fluid dynamics scenarios, such as the 3D Taylor-Green Vortex, to measure how effectively a system manages high-order finite-difference computations. Researchers and engineers utilize this benchmark to evaluate and compare the performance of different hardware configurations and computational setups in fluid dynamics simulations. Results are shown in the table below.

| Weight (DRAM) | Weight (CXL) | Execution time (s) | Speedup |
|---|---|---|---|
| 1 | 0 | 196 | 1.00 |
| 2 | 1 | 221 | 0.89 |
| 5 | 2 | 157 | 1.25 |
| 3 | 1 | 159 | 1.24 |

The benchmark has seen 25% improvement with DRAM: CXL = 5:2

**POT3D** - The Pot3D benchmark is a computational performance benchmark that simulates the 3D Poisson equation, often used to measure the performance of processors and systems in handling scientific and engineering workloads. This benchmark calculates electrostatic potentials within a 3D space, which is important in fields like molecular dynamics and computational physics. Results are shown in the table below.

| Weight (DRAM) | Weight (CXL) | Execution time(s) | Speedup |
|---|---|---|---|
| 1 | 0 | 687 | 1.00 |
| 2 | 1 | 562 | 1.22 |
| 5 | 2 | 539 | 1.27 |
| 3 | 1 | 552 | 1.24 |

The POT3D workload has demonstrated a 27% improvement with a DRAM to CXL ratio of 5:2.

*D. Putting it All Together*

Figure 3 below presents a comprehensive summary of the performance improvements observed across various HPC and AI workloads. These gains range from a 1.17x to a remarkable 1.30x enhancement, illustrating the effectiveness of integrating DDR5-6400 memory with CXL technology. By carefully calibrating the balance between DRAM and CXL memory allocations, an optimized execution configuration can be found for demanding computational tasks. For HPC and AI workloads, the geometric mean of performance speedups across all those workloads is 24%.

A notable example of these performance gains is the POT3D workload, a high-performance computing (HPC) application. The improvements in memory bandwidth and latency reduction have translated into a faster execution of complex simulations, highlighting the transformative impact of CXL memory expansion in HPC environments.

On the artificial intelligence (AI) front, the FAISS benchmark serves as a prime example. FAISS, an AI workload focused on similarity search, has shown a remarkable 23% improvement with the optimized DRAM:CXL ratio of 2:1. This gain is a testament to the enhanced memory bandwidth and performance scalability that CXL technology brings to AI applications. By leveraging the combined capabilities of DDR5-6400 and CXL-based memory expansion modules, FAISS can manage larger datasets and perform more efficient searches, thereby accelerating the overall AI processing pipeline.

V. CONCLUSION

The experimental results presented in this paper demonstrate that Micron's CZ122 CXL memory modules used in software level ratio based weighted interleave configuration significantly enhance memory bandwidth for HPC and AI workloads when used on systems with Intel's 6th Generation Xeon processors.

Key takeaways from this study include:
- Significant improvements in system performance with the combination of CXL based memory expansion and native DDR5-6400 memory due to bandwidth improvements.
- The optimization of the DRAM:CXL ratios as a critical factor in achieving these performance gains.
- The potential for CXL technology to drastically elevate the capabilities of high-performance computing and artificial intelligence applications.

The findings in this paper underscore the potential of CXL to significantly improve system efficiency and performance in demanding applications. Future research and development efforts should continue to explore and refine this integration, paving the way for even greater innovations in hybrid memory systems to meet the increasing computing demands for HPC and AI workloads.

### ACKNOWLEDGEMENTS

Thanks to the team members at Micron Technology: Eishan Mirakhur and Venkata Ravi Shankar Jonnalagadda for their contributions to the early discussions on weighted interleaving work and experiments.